# Evidence for Nitrogen Gas Surface Doping of the $Bi_2Se_3$ Topological Insulator

Michael Gottschalk,[1] Mal-Soon Lee,[2] Eric Goodwin,[1] Camille Mikolas,[1] Thomas Chasapis,[3,4] Duck Young Chung,[4] Mercouri G. Kanatzidis,[3,4] S.D. Mahanti,[1] Stuart Tessmer[1]

[1]Department of Physics and Astronomy, Michigan State University, East Lansing, Michigan 48824, USA
[2]Pacific Northwest National Laboratory, Richland, WA, 99354
[3]Department of Chemistry, Northwestern University, Evanston, Illinois 60208, USA
[4]Argonne National Laboratory, Argonne, Illinois 60439, USA

## Abstract

Using scanning tunneling spectroscopy we have studied the effects of nitrogen gas exposure on the bismuth selenide density of states. We observe a shift in the Dirac point which is qualitatively consistent with theoretical modeling of nitrogen binding to selenium vacancies. In carefully controlled measurements, $Bi_2Se_3$ crystals were initially cleaved in a helium gas environment and then exposed to a 22 SCFH flow of ultra-high purity $N_2$ gas. We observe a resulting change in the spectral curves, with the exposure effect saturating after approximately 50 minutes, ultimately bringing the Dirac point about 50 meV closer to the Fermi level. These results are compared to density functional theoretical calculations, which support a picture of $N_2$ molecules physisorbing near Se vacancies and dissociating into individual N atoms which then bind strongly to Se vacancies. In this interpretation, the binding of the N atom to a Se vacancy site removes the surface defect state created by the vacancy and changes the position of the Fermi energy with respect to the Dirac point.

## Introduction

Since their discovery, topological insulators have become intensely-studied materials for their potential applications in spintronics and topological quantum computation.[1,2] Three-dimensional topological insulators (TIs) are materials which are insulating in the bulk with a conducting surface state protected by time-reversal symmetry, making it highly robust against non-magnetic defects.[3] This topological surface state is characterized by a linear dispersion relation known as the Dirac cone which comes to a minimum, the Dirac point.[4] The Dirac point of a TI does not necessarily lie at the Fermi level, and tuning the location of the Dirac point is critical for many studies and proposed applications of the topological surface states, including superconductor-TI systems.[4–6]

$Bi_2Se_3$ is a topological insulator commonly used for research applications, as its quintuple-layer structure makes it easy to cleave with Scotch tape, and its single Dirac cone can be well resolved in tunneling spectroscopy experiments.[7–9] The Dirac point for a $Bi_2Se_3$ crystal without defects is at the Fermi level;[4] however, $Bi_2Se_3$ crystals typically feature Se vacancy defects that act as n-type dopants.[7,10,11] This leads to the occupation of the topological surface states (TSS) by the dopants and, as a result, the Dirac point for $Bi_2Se_3$ typically lies between 100 and 300 meV below the Fermi level.[7,12,13] Furthermore, the surface Se vacancies give rise to the surface defect states which can have detrimental effects on the transport properties of TSS.[10] Several studies have looked at the effects of gas exposure of $Bi_2Se_3$.[11,14,15] These studies find $O_2$, $H_2O$, and $NO_2$ to react strongly with the $Bi_2Se_3$ surface. However, an exposure effect has not, as of yet, been reported for $N_2$.

In order to understand the nature of the surface states and the effects of nitrogen exposure, we present tunneling conductance measurements acquired with a low-temperature scanning tunneling microscope (STM). We studied the density of states of $Bi_2Se_3$ crystals cleaved in $N_2$ and He gas environments, and our results show a reproducible shift in the density of states, including the Dirac point, towards the Fermi level for $N_2$-cleaved samples. We interpret these experimental results in the context of our theoretical studies of the interaction between He atom, N atom, and $N_2$ molecule with the surface of the topological insulator $Bi_2Se_3$ with and without Se vacancy. We also point out the possible role of surface states associated with the Dirac cones in dissociating the $N_2$ molecule into N atoms, which can then bind to the surface at the Se vacancy sites.

## Experimental Methods

Single crystals of $Bi_2Se_3$ samples used for this study were grown by a slow directional solidification technique. The crystals are mounted to a sample holder and placed in a cleaving chamber purged with ultra-high purity He gas (99.999%). The cleaving chamber is purged for sufficient time so that the concentration of purging gas is estimated to be 99.999%, ignoring possible outgassing; the crystal is then cleaved with Scotch tape. The sample is directly transferred to an integrated cryogenic STM without any exposure to atmosphere. For the cryogenic measurements for this study, the STM sample space is then evacuated to $10^{-6}$ torr and cooled to 77 K with liquid nitrogen. Throughout this process, the cleaved sample surface is in the overpressured environment for up to one hour. Differential tunneling conductance spectra were acquired in several different areas on multiple samples which consistently show the intrinsic n-type spectra. Figure 1(b) shows a representative spectrum taken in this manner.

To resolve the effects of nitrogen exposure in real time, we performed a series of similar measurements at room temperature. The procedure again begins with cleaving the sample in He gas and transferring the sample to the STM following the same purity measures described above. Next, we acquire tunneling conductance spectra at a single location while the system remains overpressured with He. Then, we introduced a 22 SCFH flow of pure $N_2$ gas and continue taking spectra at this location for extended periods of time by keeping the tip in tunneling range at the same location. Final spectra were taken the next day to see any possible effects of prolonged exposure. Figure 2 displays a summary of the results following this procedure.

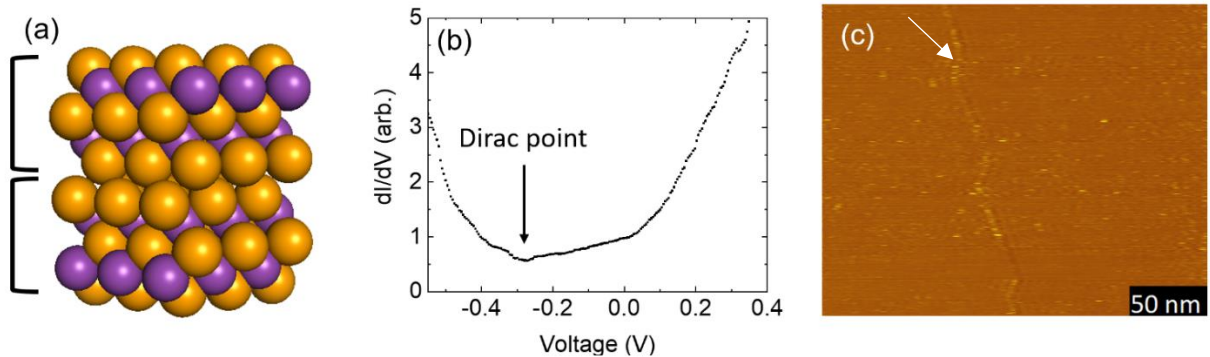

*Figure 1: (a) The quintuple layer structure of $Bi_2Se_3$. Selenium atoms comprise the surface of the quintuple layer; here two quintuple layers are shown, distinguished by the brackets. The common selenium vacancies are not shown. Color code of circles: purple – Bi, orange – Se. (b) An average of 50 tunneling conductance spectra taken on $Bi_2Se_3$ at 77K in vacuum with the tunneling barrier established by a -0.6 V bias applied to the sample and a tunneling current set point of 1nA. The valence band ends at approximately 400 meV below the Fermi level, and the conduction band begins close to the Fermi level. The Dirac point is close to 260 meV below the Fermi level, consistent with experiments on samples cleaved in vacuum. (c) A 250 x 250 nm STM topographic image showing a grain boundary feature, but otherwise featureless surface.*

## Experimental Observations

Figure 1(b) shows the baseline spectra expected for $Bi_2Se_3$. We can clearly identify the density of states steeply rising at the valence and conduction bands. As is typical for the intrinsically n-type doped $Bi_2Se_3$, the conduction band rise is close to the Fermi level, and the Dirac point, identified with the minimum of the curve, is approximately 260 meV below the Fermi level. These results are consistent with experiments on samples cleaved in vacuum and show that our purged and over-pressurized He environment is sufficiently clean to be consistent with the vacuum results for time scales up to one hour. Figure 1(c) shows a typical STM topography of the surface; although the image does not show atomic resolution, a grain boundary feature is clearly resolved. Moreover, we used atomic force microscopy (AFM) to characterize the sample surface and found an RMS roughness of 30 +/- 5 pm, consistent with studies of high-quality crystals $Bi_2Se_3$.[11]

Figure 2(a) compares our observations for helium-cleaved and nitrogen-exposed samples at room temperature. Spectra taken before the nitrogen exposure began are consistent with our results shown in Figure 1(b). After the nitrogen exposure began, a slight positive shift and lifting of the Dirac point is observed as shown in Figure 2(a). Both trends are qualitatively consistent with ab initio calculations reported below. Figure 2(b) shows tunneling current spectra as a function of tip-sample separation before and after exposure, as we expect the $N_2$ exposure to have no significant effect on the tunneling barrier, this demonstrates the stability of our tip and insensitivity to other sources of contamination; this is discussed in greater detail below. We also observe enhanced bulk contributions at the Dirac point, showing non-zero DOS. This is consistent with our density functional theory calculations in Figure 3(e) as well as our previous study.[16] We tracked the Dirac point energies, taken as the minima of the spectra with an estimated uncertainty of +/- 10 meV calculated from the scatter in the data, at the same location before introducing $N_2$ gas. The average Dirac point energy before introducing $N_2$ gas is -265 meV. The Dirac point energy is tracked as a function of time in Figure 2(c) before and after the introduction of $N_2$ gas with t=0 corresponding to when the $N_2$ gas flow began. Figure 2(c) shows a clear trend of a positive shifting of the Dirac point energy with increasing nitrogen exposure.

After 50 minutes, the Dirac point energy settles at -212 meV for a total shift of approximately 50 meV. For comparison, the last point in Figure 2(c) is the average of spectra acquired at a much greater time, 20 hours later. Here, we observe the Dirac point shifting down to -221 meV. This negative shift could possibly be attributed to water vapor outgassing from the walls of the chamber.[17] The effect on the band edges is not as pronounced; we observe evidence of a shift of approximately 10 meV that is not well-resolved in this data set.

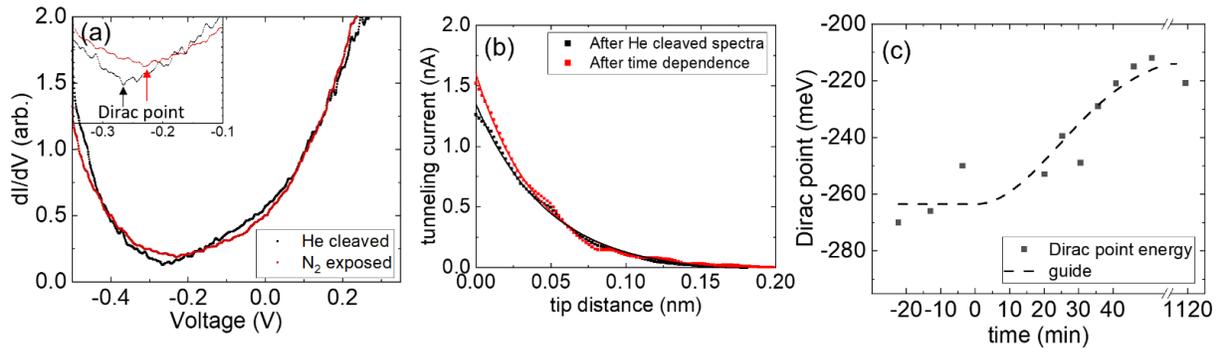

*Figure 2: (a) Representative tunneling conductance spectra acquired on $Bi_2Se_3$ cleaved in He (black) and exposed to $N_2$ (red) for 35 minutes. These spectra are each an average of 50 curves taken at room temperature at the same sample location with a tunneling barrier established by a -0.5 V bias applied to the sample and a tunneling current set point of 0.3nA with the tunneling conductance normalized to the value at 0.1 V. The inset is a smaller scale focused on the Dirac points of the two spectra. (b) Tunneling current vs. tip distance from the initial set point measurements demonstrating a robust tunneling barrier throughout the experiment. Both tunneling barriers are calculated to be greater than 5 eV, consistent with the work function of the material. The exponential decay fits are solid lines on the plot. (c) A compilation of identified Dirac point energies taken continuously at the same spot. The crystal was first cleaved in He and spectra were taken at a single location on the surface. t=0 is set when the $N_2$ gas was introduced to the system. Data taken before the scale break are extracted from averages of 50 spectra. After the scale break, we show a final point representing an average of 800 spectra taken the next day. The dashed line is a guide to the eye.*

Apparent shifts in tunneling conductance spectra can also be caused by contamination effects compromising the vacuum tunneling junction. The resulting reduction in the tunneling barrier energy can give enhancements in the conductance which are symmetric with respect to the voltage polarity. In the $Bi_2Se_3$ system, this effect may result in an apparent shift in the Dirac point. To account for this possibility, we measured the tunneling current as a function of tip distance before the first spectrum and after the last spectrum in Figure 2(c). We can see in Figure 2(b) that there is a robust tunneling barrier of at least 5 eV throughout the experiment. Hence, we can discount tunneling barrier effects as the cause of the shift. In some measurements which are not time resolved, we have observed larger shifts of approximately 150 to 250 meV after exposure to $N_2$ gas. These variations in the magnitude of the effect may arise from sample-to-sample differences in the density of selenium vacancy.

## Theoretical Calculations

Utilizing density functional theory (DFT) and the projector augmented wave method,[18] we calculated the surface local density of states for a 6 quintuple-layer (QL) $Bi_2Se_3$ slab with a 3x3 supercell in the basal (xy) plane. The slab was prepared with and without a selenium vacancy on the surface Se layer, and different atoms and molecules (loosely referred to as dopants) were iteratively brought closer to the surface while monitoring the total energy (E) of the system. Several dopants were investigated theoretically. Here we report the results for $N_2$, N and He. For the DFT calculations we used the projector augmented wave (PAW) methods[18] and the Perdew-Burke-Ernzerhof (PBE) generalized gradient corrected exchange-correlation functional[19] as implemented in the VASP package.[20] Since Bi is a heavy atom we included scalar relativistic effects and spin-orbit interaction (which are essential to get the Dirac cone states) have been included using a second variational approximation[21] in our calculations. The details of our theoretical calculations will be reported in a forthcoming paper.

Optimized structures of He/$N_2$/N on the surface with Se vacancy reveal that E is minimized when either He atom or a $N_2$ molecule is several angstroms above the surface. We define the adsorption energy $\Delta E$ as the difference of the minimum energy of the combined system from the energy of the surface plus the energy of the free dopant. A negative $\Delta E$ implies the dopant is bound to the surface. We find, for He $\Delta E$=-0.02 eV when He is above the Se vacancy and He atom does not bind with a clean surface (weak van der Waal's interaction is not adequately incorporated in our theoretical calculations). For $N_2$ molecule we find $\Delta E$=-0.09 (-0.36) eV without(with) a Se vacancy. These numbers suggest that He atom and $N_2$ molecule are more or less physisorbed on the surface with attraction to the Se vacancy, particularly by the $N_2$ molecule. In contrast, we find that $\Delta E$ is the lowest when a N atom is brought to the selenium vacancy and the corresponding adsorption energies are $\Delta E$=-1.78 (-4.82) eV, indicating strong chemical binding between the N atom and vacancy (actually the neighboring Bi atoms of a Se vacancy). It is interesting to note that a neutral N atom binds rather strongly with the surface of $Bi_2Se_3$.

For a quantitative description of the position of the dopants with respect to the surface, we define a distance $\Delta z$ (X) =z(X)-z(Se1), where z(X) is the z coordinate of X and Se1 is the surface Se atom which is removed to create a Se vacancy. We find $\Delta z$(N)=-0.77 Å, $\Delta z$(He)=2.01 Å, and $\Delta z$($N_2$)=2.06 Å, 3,16 Å. For the $N_2$ molecule we have two numbers one for each N atom, the molecular axis is nearly perpendicular to the surface. The distance between the N atoms is 1.1 Å which is very close to the bond length of a $N_2$ molecule, 1.098 Å. These distances also suggest that He and $N_2$ molecule are physisorbed whereas N atom is chemisorbed.

We calculated the density of states (DOS) of He/N adsorbed $Bi_2Se_3$ in the presence of Se vacancy. Figures 3(a) and 3(b) show the total DOS and contributions of adsorbed atom as well as Bi/Se atoms from different QL representing bulk (middle layers) and surface (top layer) in a energy range between -2 and 2 eV. We also show DOS corresponding to the energy scale of tunneling measurements for the comparison, see Figures 3(c) and 3(d). We find a fundamental difference between the position of the chemical potential (assumed to be the zero of energy) and the minimum of DOS in the gap region associated with the DP. For He, DP is ~250 meV below the chemical potential, see Figure 3(a). Also one sees a small peak in the DOS near the chemical potential (Figure 3(c)), which is ascribed to the Se1 vacancy induced surface state.[22] Clearly, states above the DP are occupied by electrons donated by the Se1 vacancy. The situation changes when one has a bound N atom as shown in Figures 3(b) and 3(d). The Se1 vacancy induced surface defect state is moved away from the gap region. The chemical potential falls below the DP by ~150 meV and even goes below the valence band maximum, indicating hole doping. The direction of the shift is consistent with the STM experiment. However, the magnitude of the shift in the calculations (400 meV) is greater than what was observed in the experiment (50 meV.) This suggests that only a fraction of the selenium vacancies bind with $N_2$ molecules.

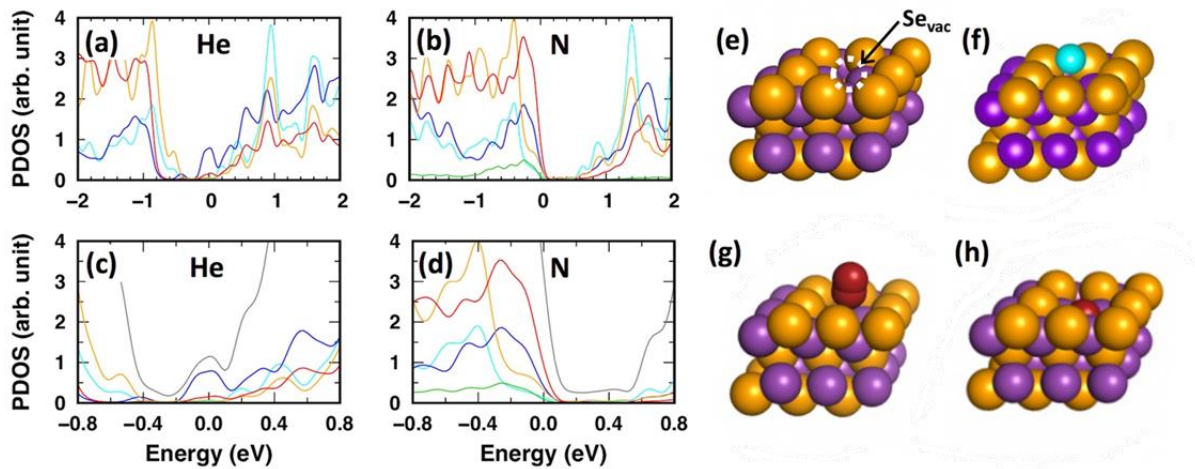

*Figure 3. (a) and (b) Calculated density of states for He and N adsorbed Bi2Se3 in the presence of Se vacancy. (c) and (d) DOS calculations corresponding to the energy scale of tunneling measurements. Color code of lines: red – surface Se, orange – bulk Se, blue – Bi surface, cyan – Bi bulk, and gray – total DOS. (e) The structure of the top quintuple layer of $Bi_2Se_3$ with a Se vacancy and (f) with He. (g) and (h) Show the minimum energy configuration for the $N_2$ molecule before dissociation and the N atom after dissociation. Color code of circles: purple – Bi, orange – Se, cyan – He, and red -N.*

In order to understand the energetics of a molecular $N_2$ adsorbing as N atoms on the $Bi_2Se_3$ surface (with and without Se vacancy) we look at the binding energy of a N atom on the $Bi_2Se_3$ surface. We find this to be 1.78 eV without a Se vacancy and 4.82eV with a Se vacancy. The nominal bond dissociation energy of a free $N_2$ molecule, that is the energy to break a $N_2$ molecule into two N atoms, is 9.79 eV.[23] Thus breaking the $N_2$ bond on a clean $Bi_2Se_3$ surface will cost nearly 6.23 eV, clearly impossible. However, if there are two Se vacancies then the energy cost to dissociate is only ~150 meV, not difficult to circumvent. Also improved calculations might reduce this energy, even making it negative. Given this energetics, we propose a possible multi-stage process where the $N_2$ molecule is first physisorbed at the vacancy site, its triple bond is weakened by charge transfer from the TSS, followed by molecular dissociation. Such a charge transfer induced weakening of the strong $N_2$ bond and molecular dissociation occurs on transition metal surfaces.[24]

## Conclusions

STM measurements provide experimental evidence that exposure to a pure $N_2$ gas environment can affect the surface density of states of $Bi_2Se_3$ with Se vacancies, shifting the Fermi level closer to the Dirac point. This is supported by consecutive tunneling conductance spectra taken while $N_2$ gas was introduced to a clean surface at room temperature. We provide evidence that the exposure effect occurs over the course of more than an hour. We interpret the effect to be due to $N_2$ molecules physisorbing near Se vacancies and dissociating into individual N atoms which bind strongly to Se vacancies. The binding of the N atom to a Se vacancy site removes the surface defect state created by the vacancy and changes the position of the Fermi energy with respect to the Dirac point. This interpretation is supported by our *ab initio* DFT based calculations; however, more theoretical study is needed for a deeper understanding of the molecular binding and dissociation process on the surface of a topological insulator with and without surface vacancies.

## Acknowledgement

Computational resource was provided by the National Energy Research Scientific Computing Center (NERSC), a U.S. Department of Energy Office of Science User Facility operated under Contract No. DE-AC02-05CH11231. At Argonne National Laboratory samples synthesis and characterization were supported by the U.S. Department of Energy, Office of Science, Basic Energy Sciences, Materials Sciences and Engineering. Computational resource was provided by the National Energy Research Scientific Computing Center (NERSC), a U.S. Department of Energy Office of Science User Facility operated under Contract No. DE-AC02-05CH11231. At Argonne National Laboratory samples synthesis and characterization were supported by the U.S. Department of Energy, Office of Science, Basic Energy Sciences, Materials Sciences and Engineering.